\documentclass[rmp,twocolumn]{revtex4}

	\usepackage{graphicx}
\usepackage{epsfig}
\newcommand{\figwidth}{3. in}
\usepackage[greek,english]{babel}
\begin{document}
\latintext
\title{When perceptual time stands still: Long stable
memory in binocular rivalry}
\author{Efstratios Manousakis}

\address{Department of Physics, Florida State University, 
Tallahassee, Florida, 32306-4350, USA and\\
Department of Physics, University of Athens, 
Panepistimioupolis, Zografos, Athens, 157 84, Greece.}

\begin{abstract}
We have carried out binocular rivalry experiments with a large number of 
subjects to obtain high quality statistics on probability distribution of
dominance duration (PDDD)  for two cases where (a) the rival stimulus is
continuously presented and (b) the rival stimulus is periodically
removed, with stimulus-on and stimulus-off intervals ${\mathrm{T}}_{on}$ and 
${\mathrm{T}}_{\mathrm{off}}$ respectively.  It is shown that the PDDD
obtained for the latter case 
can be reproduced to a reasonable degree of approximation
by simply using the PDDD of part (a)
and slicing it at pieces of time extent ${\mathrm{T}}_{\mathrm{on}}$ and by introducing 
intervals of length ${\mathrm{T}}_{\mathrm{off}}$ between the on-intervals where
the PDDD is set to zero.   This suggests that the variables representing 
the perceptual state
do not change significantly during long blank intervals.
We argue that these findings impose challenges to theoretical
models which aim at describing visual perception. 

\end{abstract}

\maketitle
\section{Introduction}
\label{introduction}
Bistable perception \cite{Leopold99} is the spontaneous alteration between two
perceptual states which occurs upon presentation of an ambiguous 
stimulus. A form of bistable perception is binocular rivalry (BR)
\cite{visualcompetetion}
where each retina is presented with a different
image dichoptically; the observer experiences perceptual rivalry
between the two different percepts which correspond to the images 
presented to the two retinas. This phenomenon can be useful in 
neuroscience for many different reasons;
for example, it permits dissociation of neural activity which corresponds to
conscious perception from that which is related to sensory stimulation
and, thus, it has been used as a tool to investigate the neural correlates of
consciousness \cite{Crick95,sterzer02,sterzer08,sterzer09,Tong06}.

Nearly half a century ago, it was discovered \cite{orbach} that intermittent 
presentation of a bistable pattern, which appears and disappears
every few seconds, decreases the frequency of perceptual changes in
bistable perception. In a more recent remarkable study \cite{stablerivalry}, 
it was  demonstrated that a periodic removal of the stimulus, which
causes bistable perception,
can lead to nearly freezing the same perception of the rival stimulus. 
Following these measurements, Chen and He \cite{Chen04}, in the case of 
binocular rivalry, allowed 
for swapping of the two rival stimuli between the two eyes
during intermittent presentations. They found that the stabilization 
occurs only when the same stimulus was presented to the same
retinal location. Furthermore, Chen and He \cite{Chen04} and Pearson and 
Clifford \cite{Pearson04} have studied the consequence of swapping 
other characteristics of the stimuli, such as color and orientation. 
These results have been interpreted that the neural processing of the
bistable stimuli involves a form of memory storage across the
blank interval \cite{Oshea,Pearson08}.

While for the case of binocular rivalry with continuous stimulus 
there are available results for the so-called probability distribution
of dominance duration (PDDD) \cite{Levelt68,Lehky95}, so far, there are
no studies of the PDDD for the case of periodically interrupted 
stimulus. One reason for the lack of such studies may be
the fact that, in order to obtain reasonably good statistics
of the PDDD in the case of interrupted stimulus, 
compared to the continuous
stimulus case, a much larger total subject test-time  is required.  
In this paper we have carried out binocular rivalry experiments 
using a large number of subjects  for high-quality statistics
to obtain the probability distribution of dominance duration (PDDD) 
for two cases where (a) the stimulus is
continuous and (b) the stimulus is periodically removed.
The definition of the PDDD for case (a) is well-known.
We define the PDDD for case (b) as the probability density for  
a perceptual change to occur at time $t$, measured from the time 
of the previous percept change, under the condition that
the stimulus on-off periodic cycle starts at precisely the time of the
previous recorded event of percept change.  

We find that the PDDD for the case where the stimulus is periodically
removed with stimulus-on and stimulus-off intervals ${\mathrm{T}}_{\mathrm{on}}$ and 
${\mathrm{T}}_{\mathrm{off}}$ respectively, can be constructed
from the PDDD which is obtained for continuously presented stimulus
by simply slicing the distribution to
pieces of time extent ${\mathrm{T}}_{\mathrm{on}}$ and introducing intervals of time extent
${\mathrm{T}}_{\mathrm{off}}$ between these slices where the PDDD is set to zero. 
Inversely, starting from the PDDD obtained
with the periodically removed stimulus and simply erasing the blank intervals
and ``sewing'' together the stimulus-on pieces of the distribution, the
PDDD for continuous stimulus can be approximately constructed.
This suggests that there is approximately no significant change of the 
state of the perceived stimulus during the blank intervals.

The paper is organized as follows. In Sec.~\ref{method} we
discuss what precisely we measured and we present the technical 
details of the method used. In Sec.~\ref{results}, we discuss the
experimental results for the cases of continuous and 
periodically interrupted stimulus. In Sec.~\ref{analysis}
we analyze our results using the results
for continuous PDDD and in Sec.~\ref{discussion} we discuss
what we believe our results mean for our understanding
of the phenomenon of binocular rivalry and for the constraints
that they imply for theories of bistable perception.
\section{Experimental method}
\label{method}

\begin{figure}[htp]
\vskip 0.3 in 
\begin{center}
\includegraphics[width=\figwidth]{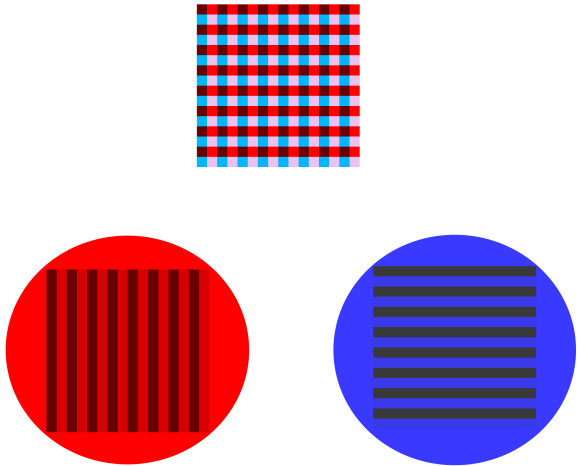}
 \caption{\label{fig1} Top: The image or stimulus which is presented to the
subjects. Bottom: The images as seen by the left and right retina when the
subjects wear glasses which have red/blue filters for the left and right
eye respectively }
\end{center}
\vskip 0.2 in 
\end{figure}
\subsection{Subjects}
A large number of subjects (53 total) participated in the
experiments reported in the present work
and they were all undergraduate students at Florida
State University (FSU). The project received prior approval
by the FSU Human Subjects Committee and the students
who agreed to participate in this study signed an appropriate
consent form. In order to achieve good statistical
quality of the probability distribution, especially for
the case of periodically removed stimulus, such a large number
of subjects was necessary.

\subsection{Apparatus and Stimuli}
A stimulus similar to the image shown as in
Fig.~\ref{fig1} (top) was projected on 
a Viewsonic Optiquest Q71 $17''$ CRT display (at $1024 \times 768$ resolution
and 60-Hz refresh rate) with 0.23 mm dot pitch  
which was connected to a computer;  the display was controlled through
a written and compiled Labview application running under the Windows operating
system. This application a) takes 
input from the mouse interrupt  when the mouse button was pressed by the 
subject, b) records the time of such an event, and 
c) resets the timer  which controls
the periodically presented stimulus to zero when the subject presses the
mouse button (see discussion below). 
Each subject was viewing the stimulus
by wearing a pair of glasses which
had a red and a blue filter for the left and right eye respectively.
The colors of the 
image, as viewed on the above described display, were carefully chosen  
such that, through the
red/blue filters used, the left or right retina would see clear 
horizontal or vertical stripes similar to those
shown on the left or the right bottom of  Fig~\ref{fig1}.
The images shown at the bottom of Fig.~\ref{fig1}, when viewed
in color (in the on-line version not the printed version of the 
paper), are similar to what an observer sees through the filters by 
closing the other eye.
The actual size of the stimulus was 4.5 cm $\times$ 4.5 cm and was set at
a distance of 100 cm from the subject's eyes ($2.6^{\circ}\times 2.6^{\circ}$
and $0.32^{\circ}$ distance between two successive horizontal or 
vertical stripes).   With the choice of contrast level made for the two
sets of stripes, we found that the average dominance duration for 
either eye was approximately the same \cite{Levelt68}.

\begin{figure}[htp]
\vskip 0.4 in 
\begin{center}
\includegraphics[width=\figwidth]{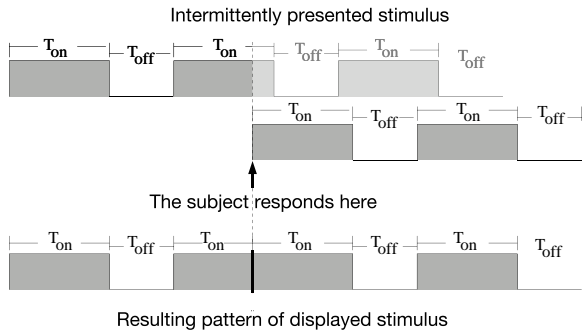}
 \caption{\label{fig2} Schematic drawing of the process
of the periodic removal of the stimulus. The gray-shaded areas
labeled as ${\mathrm{T}}_{\mathrm{on}}$ denote the periods where the stimulus is 
presented to the subject and the blank areas labeled
as ${\mathrm{T}}_{\mathrm{off}}$ denote the time intervals where the stimulus is removed.
At an instant of time $t$, indicated by the arrow, the subject responds
to a perceptual change by pressing the button. Then
the sequence of on-off periods is substituted by a
new sequence shown in the second row where the beginning of 
the periodic process starts at time $t$. The resulting pattern
of stimulus on-off periods is shown in the bottom part of the
above figure.}
\end{center}
\vskip 0.2 in 
\end{figure}

The stimulus was periodically presented for
a time interval ${\mathrm{T}}_{\mathrm{on}}$ and 
removed for a time interval ${\mathrm{T}}_{\mathrm{off}}$,
as long as there was no indication that the subject perceived a change.
Fig.~\ref{fig2} (top) schematically presents the ``on''(shaded area) and 
``off'' (blank) periods  in which the stimulus is presented
in a periodic fashion. 
The subjects were instructed to press a computer mouse button
when they perceive a change in the percept,
i.e., from horizontal to vertical stripes or vise versa. When the subject
presses the button an event is registered and at the same
time, the clock that measures the length of the presentation of the 
stimulus is reset at the beginning of the
periodic cycle. This is shown in  Fig.~\ref{fig2}(bottom) where,
at the indicated instant $t$ in time, the observer has caused
a recording of an event; then the stimulus stays on for
an interval ${\mathrm{T}}_{\mathrm{on}}$ starting from the moment $t$ and, if no other
event registers in the mean time, the stimulus goes off 
for an interval ${\mathrm{T}}_{\mathrm{off}}$ and so on
until the interrupt connected to the mouse registers a new event.
This was achieved using Labview programming in which the
interrupt connected to the mouse is coupled to the loop which
causes the periodic ``on-off'' cycle. When the interrupt picks up
the signal from the mouse input, it resets the loop which produces the periodic
cycle. At the end of the procedure a file containing the time series 
of the events $t_0$,$t_1$,$t_2$,...,$t_n$,$t_{n+1}$,..., for any given
subject is stored on a specified directory labeled by the
observer's assigned identification number.

\subsection{Procedure}

Before the actual
experiment, each observer was asked to wear the glasses and was presented
with the image of Fig.~\ref{fig1} (top) to make sure that he 
can experience rivalry. Then, another quick test was carried out 
where the observer 
was presented with the same stimulus but, this time, it was 
periodically removed and the observer was also asked to
press the button when he perceived a change from the image with the
stripes being horizontal (blue) to an image with vertical stripes (red) 
and vice  versa. The subject was told to pay attention, when the image
disappears and comes back, whether or not the percept comes back  different 
and if so to press the button if not to wait until it changes. 
Then the subject was informed that with his/her permission three actual 
experiments would be carried out next: in one, the stimulus would be
continuous, in the second and third, the stimulus would be periodically 
removed.
The subjects were also informed that the duration of each of these three
experiments would be approximately 10 minutes. Because some subjects either
had limited time available to devote to the experiments or because they 
experienced some degree of 
fatigue at the end of the first experiment (continuous stimulus) or the second 
experiment they were not asked to participate in all three cases planned. 
As a result in the case
of continuous stimulus, which was carried out first, all
53 observers participated in the experiments.
In the other two cases with the periodically
removed stimulus, only 42 observers participated in one case 
(with ${\mathrm{T}}_{\mathrm{on}}=1.1$ sec and 
${\mathrm{T}}_{\mathrm{off}}=3.1$ sec) and only 40 observers
participated in the other case (with ${\mathrm{T}}_{\mathrm{on}}=2.2$ 
sec and ${\mathrm{T}}_{\mathrm{off}}=3.1$ sec).

In producing the histograms, the time
series of the events $t_0$, $t_1$,
$t_2$,...,$t_n$, $t_{n+1}$,..., for any given
subject, is converted to a series 
$\delta t_1=t_1-t_0$, $\delta t_2=t_2-t_1$,
..., $\delta t_n =t_{n+1}-t_n$, ... of 
elapsed times between events and the histogram of these 
intervals is shown in Fig.~\ref{fig3}, 
Fig.~\ref{fig4}. 

The case of continues stimulus is obtained when 
${\mathrm{T}}_{\mathrm{on}}$ is set to be
longer than the time that the entire experiment lasts.

\section{Results}
\label{results}
\subsection{Continuous stimulus}

\begin{figure}[htp]
\begin{center}
\includegraphics[width=\figwidth]{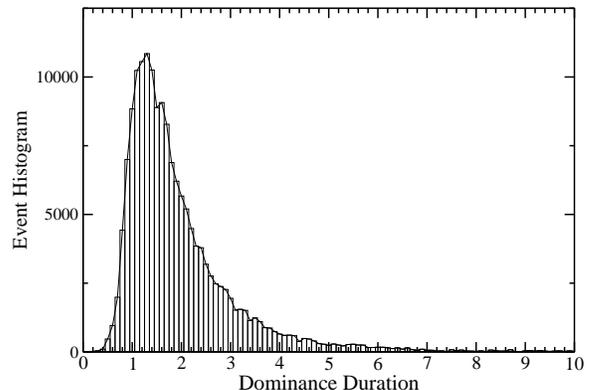}
 \caption{\label{fig3} The probability distribution of 
dominance duration obtained obtained by averaging across subjects. }
\end{center}
\end{figure}

The stimulus shown in Fig.~\ref{fig1} was continuously 
presented to 53 subjects through the two different filters for each eye as
discussed above for approximately 10 minutes per subject. 
A total number of 16435 responses (events) were recorded. 
In Fig.~\ref{fig3} we present the histogram of the event frequency. 
The calculated average value of dominance duration
is $2.05 \pm 0.02$ sec.  
\begin{figure}
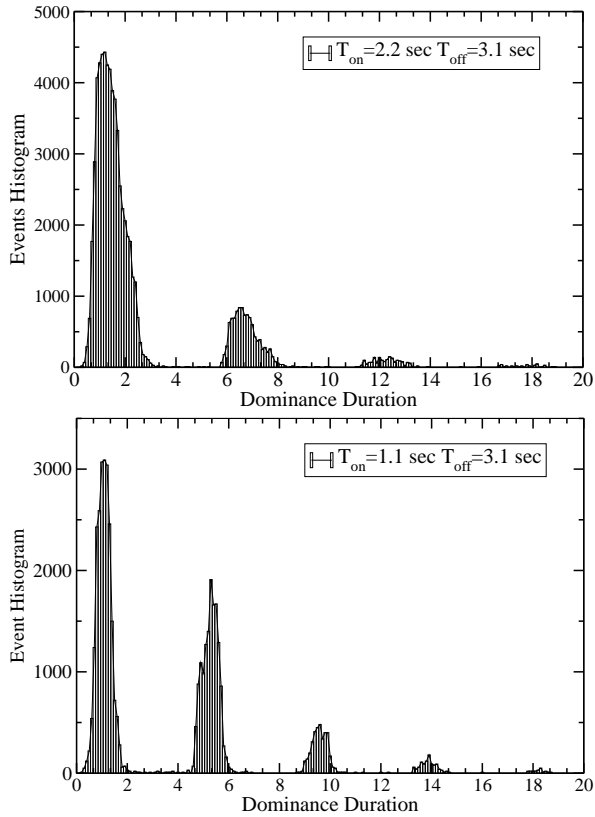

\begin{center}
\begin{tabular}{cc}
\epsfig{file=Fig4a.eps,width=0.9\linewidth,clip=}\\ 
\epsfig{file=Fig4b.eps,width=0.9 \linewidth,clip=}
\end{tabular}
 \caption{\label{fig4} 
(Top) The measured probability distribution of dominance duration (PDDD) 
averaged
over 42 subjects for  ${\mathrm{T}}_{\mathrm{on}}=2.2$ sec 
and ${\mathrm{T}}_{\mathrm{off}}=3.1$ sec.
(Bottom) The measured PDDD averaged
over 40 subjects for the case of ${\mathrm{T}}_{\mathrm{on}}=1.1$ sec and 
${\mathrm{T}}_{\mathrm{off}}=3.1$ sec.}
\end{center}
\end{figure}
\label{analysis}
\begin{figure}
\begin{center}
\includegraphics[width=\figwidth]{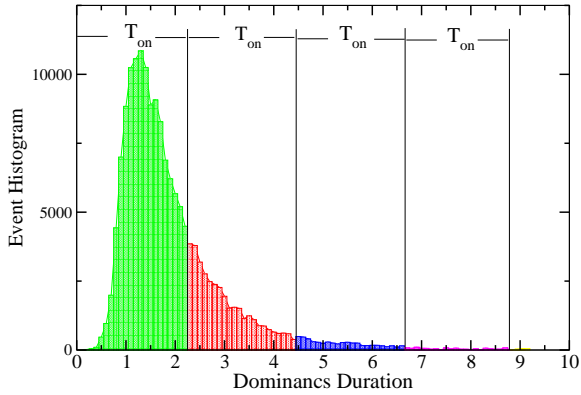}
 \caption{\label{fig5} 
Illustration of how to obtain the probability distribution of
dominance duration (PDDD) for the case where the stimulus is 
periodically interrupted from the PDDD obtained for the 
case of continuous stimulus presentation. The PDDD for continuous
stimulus is sliced at segments of duration ${\mathrm{T}}_{\mathrm{on}}$ and, 
then, they
are separated by blank intervals of duration ${\mathrm{T}}_{\mathrm{off}}$. Thus,
we obtain the PDDD for interrupted stimulus shown in Fig.~\ref{fig6}.}
\end{center}
\end{figure}

\begin{figure}
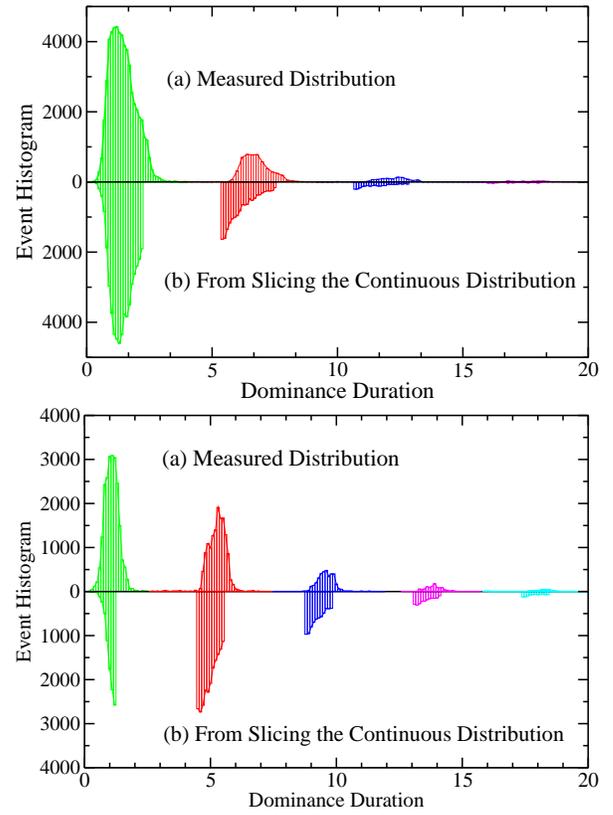

\begin{center}
\begin{tabular}{cc}
\epsfig{file=Fig6a.eps,width=0.9\linewidth,clip=}\\ 
\epsfig{file=Fig6b.eps,width=0.9 \linewidth,clip=}
\end{tabular}
 \caption{\label{fig6} 
The measured PDDD
for the case of ${\mathrm{T}}_{\mathrm{on}}=2.2$ sec, 
${\mathrm{T}}_{\mathrm{off}}=3.1$ sec  (Top) and for 
the case of ${\mathrm{T}}_{\mathrm{on}}=1.1$ sec, 
${\mathrm{T}}_{\mathrm{off}}=3.1$ sec  (Bottom) is compared to
the distribution shown as an inverted curve (part b), which is
obtained by slicing the $C(t)$, i.e., the one for continuous stimulus.}
\end{center}
\end{figure}
\subsection{Periodic removal of stimulus}
Here, we will present the results of our experiments where the stimulus was
periodically removed as discussed in Sec.~\ref{method}.
The intervals $\delta t_i$ of elapsed time between successive responses
are the dominance durations
of either percept given the constraint that, starting from the onset 
of a perceptual change, the stimulus is kept on
for a time ${\mathrm{T}}_{\mathrm{on}}$ and, then,
the stimulus is turned off for an amount of time
${\mathrm{T}}_{\mathrm{off}}$, etc., until the next perceptual change occurs
at which point in time, the clock is reset to zero and 
the onset of a new periodic cycle starts.
Data were collected for 40 observers for the case where 
${\mathrm{T}}_{\mathrm{on}}=2.2$ sec and ${\mathrm{T}}_{\mathrm{off}}=3.1$ 
sec, and the histogram 
of the $\delta t_i$ is shown in Fig.~\ref{fig4}(Top). 
In Fig.~\ref{fig4}(Bottom) the results are presented
for the case of 42 observers when  ${\mathrm{T}}_{\mathrm{on}}=1.1$ sec and 
${\mathrm{T}}_{\mathrm{off}}=3.1$ sec.

\section{Analysis of the Results}

As illustrated in Fig.~\ref{fig5}, the experimental PDDD obtained for 
continuous stimulus (after proper normalization to the same number of
events) is  sliced in parts of width equal to 
${\mathrm{T}}_{\mathrm{on}}$ and, then,
these parts are separated by intervals of time extent 
${\mathrm{T}}_{\mathrm{off}}$ as shown in
the inverted curves of Fig.~\ref{fig6}.
Namely, a discontinuous PDDD $D(t)$, to be compared with 
the case of periodically 
interrupted stimulus, is constructed using the PDDD $C(t)$ for
the case of continuous stimulus presentation, as follows:
\begin{eqnarray}
D(t)  =  \left ( \begin{array}{cc} C(t- n T_{off}),
 &  n T < t < n T + T_{on} \\
0, & otherwise \end{array} \right )
\label{interrupted}
\end{eqnarray}
where $n = [ {t \over {\mathrm{T}}}]$ (the brackets stand for the integer part)
and ${\mathrm{T}}={\mathrm{T}}_{\mathrm{on}}+
{\mathrm{T}}_{\mathrm{off}}$, the period of the on and off process.

Notice in Fig.\ref{fig5}
that several of the main features of the PDDD for the case of 
interrupted stimulus are reasonably well reproduced using the PDDD 
for continuous stimulus by this procedure.
For example, the number of on-intervals where there is a significant
probability as well as the distribution of the probability in
each of the on-intervals is similar. Furthermore, when 
the value of ${\mathrm{T}}_{\mathrm{on}}$ is smaller the number of on-interval
with significant probability increase. 

Some of the differences between the sliced-up continuous distribution 
(Eq.~\ref{interrupted})
and the experimental PDDD for the intermittent stimulus can be reconciled
as follows.
As it may be noticed, there are gaps of events in the off intervals, however,
there are also events which fall in the gaps. The main reasons
for this are \\

\noindent
(a) the observer's finite response time and \\
\noindent
(b) when the 
stimulus reappears, the observer delays to give his/her response 
mainly because it takes time to make a decision on
whether or not the percept is the same with the one perceived at the end
of the previous stimulus presentation.\\

\begin{figure}
\begin{center}
\includegraphics[width=\figwidth]{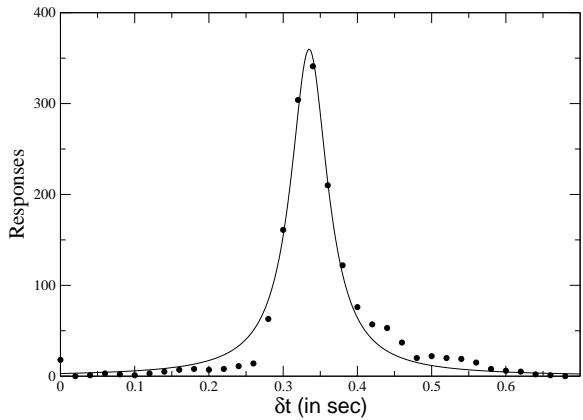}
 \caption{\label{fig7} 
The measured probability distribution of time-delay $\delta t$
of the reported time (due to finite human reaction)
from the time (as measured by a photo-diode) where the stimulus actually  
disappears. The solid line is a Lorentzian fit.} 
\end{center}
\end{figure}
\begin{figure}
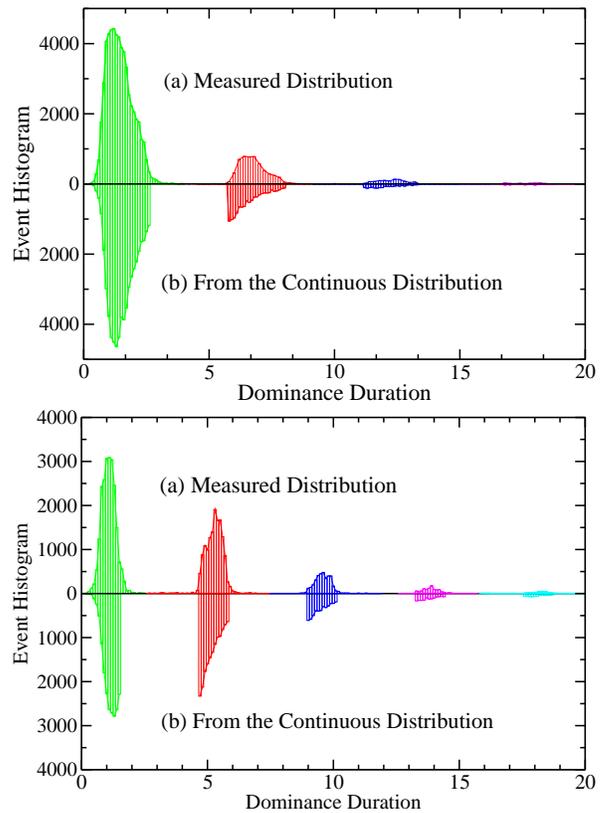

\begin{center}
\begin{tabular}{cc}
\epsfig{file=Fig8a.eps,width=0.9\linewidth,clip=}\\ 
\epsfig{file=Fig8b.eps,width=0.9 \linewidth,clip=}
\end{tabular}
 \caption{\label{fig8} 
The measured PDDD for the case of ${\mathrm{T}}_{\mathrm{on}}=2.2$ sec,
${\mathrm{T}}_{\mathrm{off}}=3.1$ sec 
(Top) for the case of ${\mathrm{T}}_{\mathrm{on}}=1.1$ sec,
${\mathrm{T}}_{\mathrm{off}}=3.1$ sec   (Bottom) sec is compared to
the distribution shown as an inverted curve (part b), which is 
obtained by slicing the $C(t)$, i.e., the 
one for continuous stimulus 
where a  $0.3$ sec is added
due to time-delay due to finite human reaction time.}
\end{center}
\end{figure}
\begin{figure}
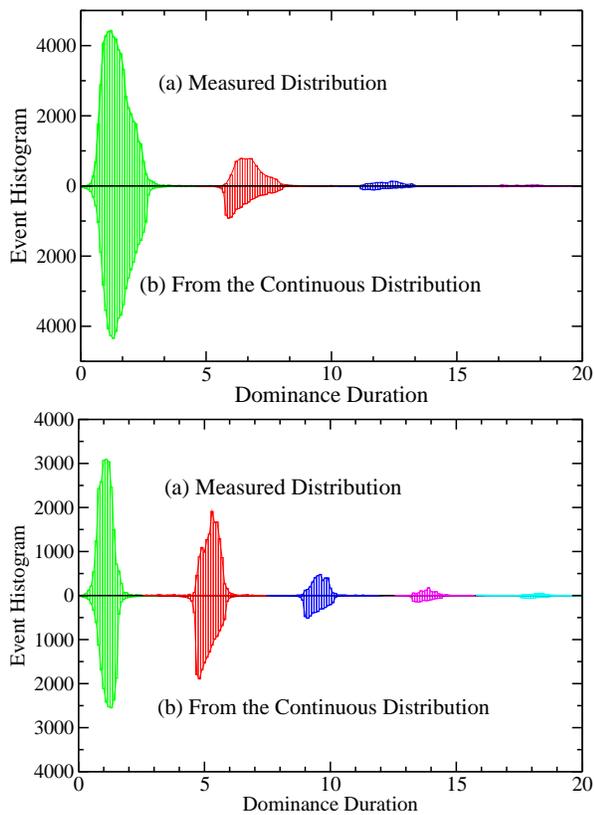

\begin{center}
\begin{tabular}{cc}
\epsfig{file=Fig9a.eps,width=0.9\linewidth,clip=}\\ 
\epsfig{file=Fig9b.eps,width=0.9 \linewidth,clip=}
\end{tabular}
 \caption{\label{fig9} 
The measured PDDD
for the case of ${\mathrm{T}}_{\mathrm{on}}=2.2$ sec, 
${\mathrm{T}}_{\mathrm{off}}=3.1$ sec (Top) and for the case 
${\mathrm{T}}_{\mathrm{on}}=2.2$ sec,
${\mathrm{T}}_{\mathrm{off}}=3.1$ sec (Bottom)  is compared to
the distribution shown as an inverted curve (part b), which is
obtained  by slicing the $C(t)$, i.e.,  the  one for continuous stimulus
where a $0.3$ sec is added
due to time-delay caused by finite human reaction time and, in addition,
by using a Lorentzian broadening of the responses with the 
independently measured width of $\epsilon=0.03$ secs.}
\end{center}
\end{figure}

In order to make sure that this is not due to an equipment delay
caused by some unknown factor, we use a photo-diode to measure 
the physical values of 
${\mathrm{T}}_{\mathrm{on}}$ and ${\mathrm{T}}_{\mathrm{off}}$.
Namely, a photo-diode connected to an oscilloscope is placed in front
of the computer monitor and, thus, the oscilloscope registers different 
voltage values depending on whether on the computer monitor  the image/stimulus
is on or off. Using the same experimental equipment, i.e., computer,
monitor and software, we presented the same stimulus with the same
two conditions of on and off times presented to the observers.
We found that both $T_{on}$ and $T_{off}$ were very close  
(within less than 10 milliseconds)
to the input values used in this study. 

Then, we carried out
a different experiment to measure the time delay 
$\delta t$ in human response from
the actual time the stimulus goes off to the time
recorded. For this experiment the same stimulus was periodically
presented and removed to observers (this time without wearing
glasses) and we asked them to respond when they saw the
stimulus disappearing. In Fig.~\ref{fig7} we present the distribution
of the time-delay measured from the actual time where the
stimulus is removed. Notice that the responses peak
at $\delta t_{\max}=0.34$ secs. The solid line is a Lorentzian
fit, i.e., to the form $C/((t-t_0)^2+\epsilon^2)$ which gives
$t_0=0.335$ secs and $\epsilon=0.0303$ secs. 

Guided by this observation we re-analyzed the PDDD data for the 
two cases studied here in a different way with the intention to 
take into consideration this delay in human reaction time.
Fig.~\ref{fig8} compare the
measured PDDD for the two cases studied here with the
PDDD obtained by slicing the 
one for continuous stimulus where  a time-delay of $\delta t = 0.3$ sec
has been included as follows in order to take into account the 
finite human reaction time. The first slice of $C(t)$ is taken at  
${\mathrm{T}}_{\mathrm{on}}+\delta t$ instead of at 
${\mathrm{T}}_{\mathrm{on}}$, and the time
at which the stimulus reappears is taken to be 
${\mathrm{T}}+\delta t$. Notice that the agreement
with the observed PDDD  has improved rather significantly.

In Fig.~\ref{fig9} the
measured PDDD for the two cases studied here is compared with the
PDDD obtained as discussed in the previous paragraph (for
Fig.~\ref{fig8} and, in addition,
the distribution is broadened by a Lorentzian
with width $\epsilon=0.03$, i.e., exactly the same width found experimentally
for the reaction-time distribution (Fig.~\ref{fig7})
Notice that the agreement
with the observed PDDD  has improved further.

Notice, however, that the
distribution obtained by just slicing $C(t)$ has a peak very close 
to the on-set of the new 
periodic cycle while the experimental peak is delayed.
This behavior is very similar in all of the on-intervals.
We believe that this is due to the fact (b) discussed above
namely, that when the interrupted stimulus reappears, the 
observer, in addition, to the standard response time, he needs time 
to make a conscious decision on whether the percept is the same
or different with the one he was experiencing when the stimulus
went off.



\section{Discussion and conclusions}
\label{discussion}

The PDDD for the case where the stimulus is periodically removed
has been measured for the first time. A large number of subjects
were used in order to obtain reasonably good statistics.

Our analysis of our results demonstrates that, to a 
reasonable approximation, the
main features of the PDDD with interrupted stimulus can be 
obtained from the PDDD which corresponds to the continuous stimulus
by simply slicing the latter and introducing no response between the
sliced intervals. Inversely,
if we put-together the pieces of the  PDDD corresponding to the case 
of interrupted stimulus we obtain the PDDD corresponding to the
case of continuous stimulus presentation.

The results of the experiments reported in the present work, 
imply that there is no significant
change and no significant decay of the memory of the perceived state 
during the stimulus
interruptions. While physical time is the parameter that follows
the physical change (evolution), we can define ``perceptual time''
as the parameter which characterizes perceptual change. This
latter time, to a large extent, halts during the intervals of the 
stimulus interruption.

Adaptation-inhibition models cannot explain the 
perceptual stabilization upon stimulus interruption. 
In order to describe perceptual stabilization of interrupted 
ambiguous stimuli, Noest et al. \cite{Noest07} have introduced a 
model in which percept-choice 
at stimulus onset differs fundamentally from percept-switching due to
inclusion of interaction between ``shunting'' adaptation and a near-threshold
neural baseline.
In order to describe the stabilization of the perceptual state 
in binocular rivalry Wilson \cite{Wilson07} has introduced a model 
in which he assumes a rapid synaptic potentiation
followed by a significantly slower depression back to the original
level. However, it is not at all obvious that the above models 
can reproduce this measured behavior of the PDDD.
In both theoretical models, the dynamical variables which represent
the perceived state evolve during the blank intervals; 
however, the present experimental results suggest
that the perceptual state is approximately unaffected by time during stimulus
interruptions (which last for several seconds) as shown in the present
experiment.
This experiment was motivated by a theoretical work of the present
author \cite{manousakis1}. The experimental results of
the present paper are in reasonable agreement with the
prediction of that theory, i.e., Eq.~\ref{interrupted}. 
It may still be premature to accept the rather radical theoretical
point of view of our previous theoretical work \cite{manousakis1}
without further investigation; however, the experimental
results of the present study are significant because they should serve 
as a benchmark by which to test
any theory of bistable perception.

\vskip 0.3 in 
\section{Acknowledgments}
I would like to thank S. Barton, J. Ryan and I. Winger for valuable 
discussions and K. Koetz for his very valuable technical
assistance and especially for development of software 
needed for these experiments.

\bibliographystyle{apalike}

\end{document}